\documentclass{article}

\usepackage{times}
\usepackage{hyperref}
\usepackage{algorithm}
\usepackage{algpseudocode}
\usepackage{tikz,pgf}
\usepackage{epsfig,amssymb,amsmath,ifthen,comment}
\usepackage{color}

\usepackage[normalem]{ulem}
\newcommand{\stkout}[1]{\ifmmode\text{\sout{\ensuremath{#1}}}\else\sout{#1}\fi}

\def\ci{\perp\!\!\!\perp}

\title{When does the ID algorithm fail?}

\newtheorem{assumption}{Assumption}
\newtheorem{proposition}{Proposition}

\DeclareMathOperator{\pa}{pa}
\DeclareMathOperator{\de}{de}

\DeclareMathOperator{\ch}{ch}
\DeclareMathOperator{\an}{an}

\DeclareMathOperator{\mb}{mb}

\author{
Ilya Shpitser\\
\texttt{ilyas@cs.jhu.edu}
}

\begin{document}

\maketitle

\begin{abstract}
The ID algorithm 
solves the problem of identification of interventional distributions of the form $p(\vec{Y} \mid \text{do}(\vec{a}))$ in graphical causal models, and has been formulated in a number of ways \cite{tian02on,shpitser06id,richardson23nested}.
The ID algorithm is sound (outputs the correct functional of the observed data distribution whenever $p(\vec{Y} \mid \text{do}(\vec{a}))$ is identified in the causal model represented by the input graph), and complete (explicitly flags as a failure 
any input $p(\vec{Y} \mid \text{do}(\vec{a}))$ whenever this distribution is not identified in the causal model represented by the input graph).

The reference \cite{shpitser06id} provides a result, the so called ``hedge criterion'' (Corollary 3), which aims to give a graphical characterization of situations when the ID algorithm fails to identify its input in terms of a structure in the input graph called the hedge.
While the ID algorithm is, indeed, a sound and complete algorithm, and the hedge structure does arise whenever the input distribution is not identified, Corollary 3 presented in \cite{shpitser06id} is incorrect as stated.
In this note, I outline the modern presentation of the ID algorithm, discuss a simple counterexample to Corollary 3, and provide a number of graphical characterizations of the ID algorithm failing to identify its input
distribution.
\end{abstract}

I first go over a few preliminaries on graphical causal models, interventional distributions, identification, and the ID algorithm.  If you are familiar with these concepts, you can safely skip to Section \ref{sec:counterexample} which discusses Corollary 3 in \cite{shpitser06id}.

\section{Graphical causal models and identification of causal effects}

Causal models use counterfactual variables to quantify the causal effect of one or more \emph{treatment} variables $\vec{A}$ on an \emph{outcome} variable of interest $Y$, and
are written as $Y(\vec{a})$, meaning `$Y$ had $\vec{A}$ been set, possibly contrary to fact, to value $\vec{a}$.'
Causal effects are generally conceptualized as comparisons, on the expectation scale, of outcomes in hypothetical randomized controlled trials where different arms are defined by counterfactual interventions on treatments $\vec{A}$.  Such interventions are denoted by the $\text{do}(\vec{a})$ operator in \cite{pearl09causality}.  For example, the average causal effect (ACE) is defined as $\mathbb{E}[Y(\vec{a})] - \mathbb{E}[Y(\vec{a}')]$.  Causal models are used to link counterfactual and factual variables, making inferences about causal parameters such as the ACE possible from observed data, if these parameters are \emph{identified}.

A popular causal model is associated with a directed acyclic graph (DAG) ${\cal G}$, known as the structural causal model (SCM) \cite{pearl09causality}, associates variables $\vec{V}$ and their causal relationships with vertices and edges in ${\cal G}$, as follows.
The SCM posits a set of noise variables $\epsilon_i$ for each $V_i \in \vec{V}$, such that $p(\epsilon_1, \ldots, \epsilon_k) = \prod_{i} p(\epsilon_i)$, and a set of unrestricted structural equations $f_i$ that map values of observed direct causes of $V_i$ (parents in the graph ${\cal G}$), written as $\pa_{\cal G}(V_i)$, and $\epsilon_i$ to values of $V_i$.  The set of functions $\{ f_i : V_i \in \vec{V} \}$ is meant to represent causal mechanisms that reliably map inputs to outputs, even if inputs were set by external intervention $\text{do}(\vec{a})$ that sets $\vec{A} \subseteq \vec{V}$ to $\vec{a}$.
The set of counterfactual variables after the $\text{do}(\vec{a})$ operation had been applied is denoted as $\vec{V}(\vec{a}) \equiv \{ V_i(\vec{a}) : V_i \in \vec{V} \}$.

\begin{figure*}
	\begin{center}
		\begin{tikzpicture}[>=stealth, node distance=1.0cm]
		\tikzstyle{format} = [draw, very thick, circle, minimum size=5.0mm,
		inner sep=0pt]
		\tikzstyle{unode} = [draw, red, very thick, circle, minimum size=1.0mm,
		inner sep=0pt]
		\tikzstyle{square} = [draw, very thick, rectangle, minimum size=4mm]
				
		\begin{scope}[xshift=0.0cm]
		\path[->, very thick]
		node[format] (a1) {$A_1$}
		node[format, right of=a1] (l) {$L$}
		node[format, right of=l] (a2) {$A_2$}
		node[format, right of=a2] (y) {$Y$}

		(a1) edge[blue] (l)
		(l) edge[blue] (a2)
		(a2) edge[blue] (y)
		(a1) edge[blue, bend left] (a2)
		(l) edge[blue, bend right] (y)
		(a1) edge[blue, bend left] (y)

		node[below of=l, yshift=0.3cm, xshift=0.5cm] (a) {$(a)$}
		;

		\end{scope}

		\begin{scope}[xshift=3.75cm]
		\path[->, very thick]
		node[format] (a1) {$A_1$}
		node[unode, right of=a1] (h1) {$H_1$}
		node[format, right of=h1] (y) {$Y$}
		node[format, above of=y] (a2) {$A_2$}
		node[unode, left of=a2] (h2) {$H_2$}
		node[format, left of=h2] (w) {$W$}

		(a1) edge[blue, bend right] (y)
		(a2) edge[blue] (y)
		(w) edge[blue] (a1)
		
		(h2) edge[red] (w)
		(h2) edge[red] (a2)

		(h1) edge[red] (w)
		(h1) edge[red] (y)
		
		node[below of=h1, yshift=0.3cm] (b) {$(b)$}

		;
		\end{scope}		
		
		\begin{scope}[xshift=6.5cm]
		\path[->, very thick]
		node[format] (a1) {$A_1$}
		node[format, right of=a1] (y) {$Y$}
		node[format, above of=y] (a2) {$A_2$}
		node[format, left of=a2] (w) {$W$}

		(a1) edge[blue] (y)
		(a2) edge[blue] (y)
		(w) edge[blue] (a1)
		
		(w) edge[<->, red] (a2)
		(w) edge[<->, red] (y)
		
		node[below of=a1, xshift=0.5cm, yshift=0.3cm] (c) {$(c)$}

		;
		\end{scope}

		\begin{scope}[xshift=8.25cm]
		\path[->, very thick]
		node[format] (a) {$A$}
		node[format, right of=a] (m) {$M$}
		node[format, right of=m] (y) {$Y$}
		node[format, above of=m] (c) {$C$}

		(c) edge[blue] (a)
		(c) edge[blue] (m)
		(c) edge[blue] (y)

		(a) edge[blue] (m)
		(m) edge[blue] (y)
		(a) edge[<->, red, bend right] (y)

		node[below of=m, yshift=0.3cm, xshift=0.0cm] (d) {$(d)$}
		;

		\end{scope}

		\begin{scope}[xshift=11.0cm]
		\path[->, very thick]
		node[format] (m) {$M$}
		node[format, above of=m] (c) {$C$}
		node[format, right of=m] (y) {$Y$}

		(c) edge[blue] (m)
		(c) edge[blue] (y)
		(m) edge[blue] (y)

		node[below of=m, yshift=0.3cm, xshift=0.5cm] (e) {$(e)$}
		;

		\end{scope}
		
		\end{tikzpicture}
	\end{center}
	\caption{
		(a) A four variable directed acyclic graph.
		(b) A simple counterexample to the hedge criterion.  In the model represented by this hidden variable directed acyclic graph (DAG), $p(Y \mid \text{do}(a_1,a_2))$ is identified, but $p(Y \mid \text{do}(a_2))$ is not, with the corresponding witnessing hedge being equal to the pair $\{ W, Y \}$ and $\{ Y, W, A_2 \}$.
		(c) A latent projection acyclic directed mixed graph (ADMG) of the hidden variable DAG in (b).
		(d) A latent projection representing the so called ``front-door'' model.
		(e) ${\cal G}_{\vec{Y}^*}$ obtained from the graph in (d) and the query $p(Y \mid \text{do}(a))$.
	}
	\label{fig:example}
\end{figure*}
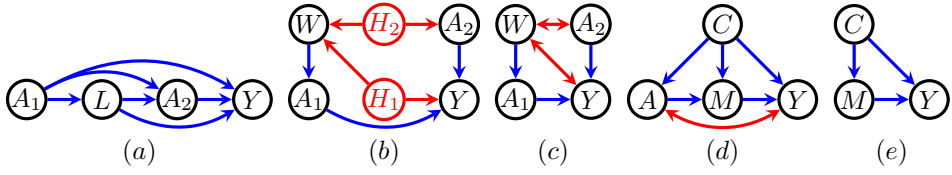

If every variable corresponding to a vertex in ${\cal G}$ (with a vertex set $\vec{V}$) representing an SCM is observed, every interventional distribution $p(\vec{Y} \mid \text{do}(\vec{a}))$ for any disjoint subsets $\vec{A},\vec{Y}$ of $\vec{V}$ is identified by the \emph{g-formula} \cite{robins86new}:
\begin{align}
p(\vec{Y} \mid \text{do}(\vec{a})) = \sum_{\vec{V} \setminus (\vec{Y} \cup \vec{A})} \prod_{V \in \vec{V} \setminus \vec{A}} p(V \mid \pa_{\cal G}(V)) \vert_{\vec{A} = \vec{a}}.
\label{eqn:g}
\end{align}
As an example, in Fig.~\ref{fig:example} (a), we have
\begin{align*}
p(Y \mid \text{do}(a_1, a_2)) = \sum_{L} p(Y \mid L, a_1, a_2) p(L_1 \mid a_1).
\end{align*}

Note that (\ref{eqn:g}) holds if $\vec{A}$ is the emptyset, implying that $p(\vec{V})$ obeys the Markov factorization of the DAG \cite{pearl88probabilistic,lauritzen96graphical}:
\begin{align*}
p(\vec{V}) = \prod_{V \in \vec{V}} p(V \mid \pa_{\cal G}(V)).
\end{align*}



\section{The ID algorithm}

In the presence of hidden variables, identification in graphical causal models becomes considerably more complicated.  Some interventional distributions become non-identified, and others are identified as much more complex functionals of the observed data distribution than the g-formula.

Pearl developed the do-calculus \cite{pearl09causality} as a general purpose deductive reasoning system for answering identification questions in causal systems, including systems with hidden variables.  Jin Tian proposed a general identification algorithm for hidden variable graphical causal models in \cite{tian02on}, which was simplified in \cite{shpitser06id} as the ID algorithm, which was further reformulated into a one line formula in \cite{richardson23nested}.  First, I briefly review the modern formulation of the ID algorithm.

Rather than taking a hidden variable DAG ${\cal G}(\vec{V} \cup \vec{H})$ as a representation of a causal model, the ID algorithm takes as input a special mixed graph called a \emph{latent projection} \cite{verma90equiv}.
Specifically, a latent projection is an acyclic directed mixed graph (ADMG), which is a mixed graph containing directed and bidirected edges and no directed cycles.

Given a DAG ${\cal G}(\vec{V} \cup \vec{H})$, where $\vec{V}$ correspond to observed variables, and $\vec{H}$ correspond to hidden variables, a latent projection ${\cal G}(\vec{V}$ contains only vertices in $\vec{V}$, and the following edges.
If $V_i \to V_j$ is an edge in ${\cal G}(\vec{V} \cup \vec{H})$, this edge also exists in ${\cal G}(\vec{V})$.  If $V_i$ is connected to $V_j$ by a directed path where all intermediate vertices are in ${\cal H}$, then an edge $V_i \to V_j$ exists in ${\cal G}(\vec{V})$.
If $V_i$ is connected to $V_j$ by a path that is not directed, does not contain any two consequtive directed edges with meeting arrowheads, and where all intermediate vertices are in ${\cal H}$, then an edge $V_i \leftrightarrow V_j$ exists in ${\cal G}(\vec{V})$.
As an example, the latent projection of a hidden variable DAG in Fig.~\ref{fig:example} (b) is shown in Fig.~\ref{fig:example} (c).

It turns out that any two hidden variable DAGs ${\cal G}_i(\vec{V} \cup \vec{H}_i)$ and ${\cal G}_j(\vec{V} \cup \vec{H}_j)$ that share the same latent projection ${\cal G}_i(\vec{V}) = {\cal G}_j(\vec{V})$ will agree on identification theory \cite{richardson23nested}.

Thus, the ID algorithm takes two inputs: an interventional distribution $p(\vec{Y} \mid \text{do}(\vec{a}))$ from a hidden variable causal model, and a latent projection ${\cal G}(\vec{V})$ representing this model.
It can be shown that the following identity holds:
\begin{align}
p(\vec{Y} \mid \text{do}(\vec{a})) = \sum_{\vec{Y}^* \setminus \vec{Y}} \prod_{\vec{D} \in {\cal D}({\cal G}_{\vec{Y}^*})} p(\vec{D} \mid \text{do}(\vec{s}_{\vec D})).
\label{eqn:line-4}
\end{align}
where
\begin{itemize}
\item $\vec{Y}^*$ is the set of elements of $\vec{V}$ that include $\vec{Y}$ and any vertex with a directed path to an element of $\vec{Y}$, such that this directed path does not intersect $\vec{A}$,
\item ${\cal G}_{\vec{Y}^*}$ is a subgraph of ${\cal G}(\vec{V})$ containing only vertices in $\vec{Y}^*$ and edges between them,
\item ${\cal D}({\cal G}_{\vec{Y}^*})$ is the set of bidirected connected sets (called \emph{districts} in \cite{richardson23nested}, and \emph{c-components} in \cite{tian02on}) in ${\cal G}_{\vec{Y}^*}$.
\item $\vec{s}_{\vec D}$ are value assignments to $\pa_{\cal G}(\vec{D}) \setminus \vec{D}$ either consistent with $\vec{a}$, or marginalized over in the outer summation.
\end{itemize}

As an example, if the ID algorithm is given $p(Y \mid \text{do}(a))$ and the graph ${\cal G}(\{ A, M, Y, C \})$ in Fig.~\ref{fig:example} (d) as input, we will have:
\begin{itemize}
\item $\vec{Y}^* = \{ Y, M, C \}$.
\item ${\cal G}_{\vec{Y}^*}$ is shown in Fig.~\ref{fig:example} (e).
\item ${\cal D}({\cal G}_{\vec{Y}^*})$ is equal to $\{ \{ Y \}, \{ M \}, \{ C \} \}$, meaning that in this case there are three relevant districts, each containing a single element.
\end{itemize}
Therefore, the decomposition of $p(Y \mid \text{do}(a))$ in this example is
\begin{align}
p(Y \mid \text{do}(a)) = \sum_{m,c} p(Y \mid \text{do}(m,c)) p(M = m \mid \text{do}(a,c)) p(c).
\end{align}

It turns out that $p(\vec{Y} \mid \text{do}(\vec{a}))$ is identified if each term in (\ref{eqn:line-4}) is identified.  To describe how the ID algorithm tries to obtain identification of each term, I borrow from the description in \cite{shpitser21proximal}.

To check identifiability of $p(\vec{D} \mid \text{do}(\vec{s}_{\vec D}))$, we will need to introduce graphs derived from ${\cal G}(\vec{V})$ that contains vertices $\vec{R}$ representing random variables, and \emph{fixed} vertices $\vec{S}$ representing intervened on variables.   Such graphs {${\cal G}(\vec{R},\vec{S})$} are called \emph{conditional ADMGs} or CADMGs, and represent {independence restrictions in} interventional distributions {$p(\vec{R} \mid \text{do}(\vec{s}))$}.

Given a CADMG ${\cal G}(\vec{R},\vec{S})$,
we say $R \in \vec{R}$ is \emph{fixable} if there \emph{does not exist} another vertex $W$ with both a directed path from $R$ to $W$ in ${\cal G}(\vec{R},\vec{S})$ (e.g. $W$ is a descendant of $R$ in ${\cal G}(\vec{R},\vec{S})$) and a bidirected path from $R$ to $W$.  Note that aside from the starting vertex $R$, and the ending vertex $W$, these two paths need not share vertices.  Given a fixable vertex $R$, a fixing operator $\phi_R({\cal G}(\vec{R},\vec{S}))$ produces a new CADMG ${\cal G}(\vec{R} \setminus \{ R \},\vec{S}\cup\vec\{ R \})$ obtained from ${\cal G}(\vec{R},\vec{S})$ by removing all edges with arrowheads into $R$. {The fixing operator provides a way of constructing the CADMG ${\cal G}(\vec{R} \setminus \{ R \},\vec{S}\cup\vec\{ R \})$ representing restrictions in $p(\vec{R} \setminus \{ R \} \mid \text{do}(\vec{s} \cup r))$, from the CADMG ${\cal G}(\vec{R},\vec{S})$, representing restrictions in $p(\vec{R} \mid \text{do}(\vec{s}))$.}  As we describe below, $R$ being fixable in ${\cal G}(\vec{R},\vec{S})$
is a graphical representation of $p(\vec{R} \setminus \{ R \} \mid \text{do}(\vec{s} \cup r))$ being identified from $p(\vec{R} \mid \text{do}(\vec{s}))$ in a particular way.

A sequence $\sigma_{\vec{J}} \equiv \langle J_1, J_2, \ldots, J_k \rangle$ of vertices in a set $\vec{J} {\subseteq \vec{R}}$ is said to be \emph{valid} in ${\cal G}(\vec{R},\vec{S})$ if it is either empty, or $J_1$ is fixable in ${\cal G}(\vec{R},\vec{S})$, and $\tau(\sigma_{\vec{J}}) \equiv \langle J_2, \ldots, J_k \rangle$ (the \emph{tail} of the sequence) is valid in $\phi_{J_1}({\cal G}(\vec{R},\vec{S}))$.  Any two distinct sequences $\sigma^1_{\vec{J}},\sigma^2_{\vec{J}}$ on the same set $\vec{J}$ valid in ${\cal G}(\vec{R},\vec{S})$ yield the same graph: $\phi_{\sigma^1_{\vec{J}}}({\cal G}(\vec{R},\vec{S})) = \phi_{\sigma^2_{\vec{J}}}({\cal G}(\vec{R},\vec{S}))$.  We will thus write $\phi_{\vec{J}}({\cal G}(\vec{R},\vec{S}))$ to denote this graph.

If a valid sequence for $\vec{V} \setminus \vec{R}$ exists in a graph ${\cal G}(\vec{V})$, the set $\vec{R}$ is called \emph{reachable}.  If, further, $\vec{R}$ is a bidirected connected set, it is called an \emph{intrinsic set}.

Each term $p(\vec{D} \mid \text{do}(\vec{s}_{\vec D}))$ in (\ref{eqn:line-4}) is identified from $p(\vec{V})$ if and only if there exists a sequence $\langle J_1, \ldots \rangle$ of elements in $\vec{J} \equiv \vec{V} \setminus \vec{D}$ valid in ${\cal G}(\vec{V})$.
{
While it might appear that checking whether elements of a set $\vec{J}$ admit a valid sequence in ${\cal G}(\vec{V})$, and finding such a sequence, if it exists, might be a computationally challenging problem, in fact this problem admits a low order polynomial time algorithm.  This is because the fixing operator applied to graphs \emph{only removes edges} and never adds edges.  Removing an edge can never prevent a vertex from being fixable if it was fixable before the edge was removed.  As a result, an algorithm looking for a valid sequence never needs to backtrack -- finding any fixable vertex among vertices yet to be fixed suffices to eventually yield a fixing sequence if it exists, while being unable to find such a vertex at some point implies such a sequence does not exist.
}

If 
a {valid} sequence exists {in ${\cal G}(\vec{V})$ for the set $\vec{J} = \{ J_1, J_2, \ldots \}$}, it implies 
{a set of} identifying assumptions {in the  causal model represented by ${\cal G}(\vec{V})$}.
Let $\vec{R}_1 = \de_{{\cal G}(\vec{V})}(J_1) \setminus \{ J_1 \}$, and $\vec{T}_1 = \vec{V} \setminus (\vec{R}_1 \cup \{ J_1 \})$.
Similarly, let
$\vec{R}_k = \de_{\phi_{\langle J_1, \ldots, J_{k-1} \rangle}({\cal G}(\vec{V}))}(J_k) \setminus \{ J_k \}$ and $\vec{T}_k = \vec{V} \setminus (\{ J_1, \ldots, J_{k-1}, J_k \} \cup \vec{R}_k)$, where $\de_{\cal G}(R_k)$ is the set of descendants of $R_k$ (including $R_k$ itself by convention) in ${\cal G}$.
{Then the graphical causal model implies the following restrictions.}
\begin{assumption}[sequential fixing ignorability]
\label{a:fix-ignore}
{\small
\begin{align}
\label{eqn:ignore-p-1}
\vec{R}_1(j_1) \ci J_1 \mid \vec{T}_1\text{ for all }j_1\\
\label{eqn:ignore-p-2}
\vec{R}_k(j_1,\ldots,j_k) \ci J_k(j_1, \ldots,j_{k-1}) \mid \vec{T}_k(j_1,\ldots,j_{k-1})\text{ for all }j_1,\ldots,j_k.
\end{align}
}
\end{assumption}
{
The identifying assumptions associated with a valid sequence of $\vec{J}$ may be viewed as a{n inductive} generalization of conditional ignorability, or sequential ignorability \cite{robins86new}.
At the point of the induction when a particular variable $J_k(j_1, \ldots, j_{k-1})$ is fixed, 
it is viewed as a ``treatment,'' while all variables in $\vec{R}_k(j_1, \ldots, j_{k-1})$ are viewed as ``outcomes,'' and all variables in $\vec{T}_k(j_1, \ldots, j_{k-1})$ are viewed as ``observed covariates.''
}

Given 
{assumption \ref{a:fix-ignore}}, we obtain identification of $p(\vec{D} \mid \text{do}(\vec{s}_{\vec D}))$ by the following inductive formula:
{\small
\begin{align}
\notag
p(\vec{V} \setminus \{ J_1 \} \mid \text{do}(j_1))
&=
\frac{
p(\vec{V} \setminus \{ J_1 \}, j_1)
}{
p(j_1 \mid \vec{T}_1)
}
=
\frac{
p(\vec{V} \setminus \{ J_1 \}, j_1)
}{
p(j_1 \mid \mb^*_{{\cal G}(\vec{V})}(J_1))
}\\
\notag
p(\vec{V} \setminus \{ J_1, \ldots, J_k \} \mid \text{do}(j_1, \ldots, j_{k}))
&=
\frac{
p(\vec{V} \setminus \{ J_1, \ldots, J_{k} \}, j_k \mid \text{do}(j_1, \ldots, j_{k-1}))
}{
p(j_k \mid \vec{T}_k, \text{do}(j_1, \ldots, j_{k-1}))
}\\
&=
\frac{
p(\vec{V} \setminus \{ J_1, \ldots, J_{k} \}, j_k \mid \text{do}(j_1, \ldots, j_{k-1}))
}{
p(j_k \mid \mb^*_{\phi_{\langle J_1, \ldots, J_{k-1} \rangle}({\cal G}(\vec{V}))}(J_k), \text{do}(j_1, \ldots, j_{k-1}))
},
\label{eqn:one-district}
\end{align}
}
where for any $J_i \in {\vec{V}} \setminus \{ J_1, \ldots, J_{i-1} \}$, $\mb^*_{{ {\cal G}^* }}(J_i)$ denotes all random vertices that are either parents of $J_i$, or that are connected to $J_i$ via collider paths (paths where all consecutive triplets have arrowheads meeting at the middle vertex) {in a CADMG ${\cal G}^*$}.
The operations on the right hand side of (\ref{eqn:one-district}) may be viewed as distributional analogues of the graphical fixing operation $\phi$,
{and are licensed by repeated applications of assumption \ref{a:fix-ignore} and consistency, or alternatively as applications of rule 2 of the potential outcomes calculus \cite{malinsky19po,rrs22volume_id}.
}

As a simple example, we illustrate how identifiability of 
{$p(Y \mid \text{do}(a))$} in Fig.~\ref{fig:example} (e)
in terms of (\ref{eqn:line-4}) and (\ref{eqn:one-district}).
If we aim to identify $p(Y \mid \text{do}(a))$ in Fig.~\ref{fig:example} (e), we note that $Y^* = \{ Y, M, C \}$, with districts in ${\cal G}_{Y^*}$ being $\{ Y \}, \{ M \}, \{ C \}$.
In fact, valid sequences exists for all sets of elements outside these districts.  Thus, we have the following derivation for the term $p(C \mid \text{do}(a,y,m))$:
{\small
\begin{align*}
p(C,A,M\mid\text{do}(y)) &= \frac{p(C,A,M,y)}{p(y \mid C,A,M)} {= p(C,A,M)}\\
p(C,A\mid\text{do}(y,m)) &= \frac{p(C,A,m \mid \text{do}(y))}{p(m \mid A,C,\text{do}(y))} = \frac{p(C,A,m)}{p(m \mid A,C)} {=p(C,A)}\\
p(C \mid \text{do}(a,y,m)) &= \frac{p(C,a \mid \text{do}(y,m))}{p(a \mid C,\text{do}(y,m))} = \frac{p(C,a)}{p(a \mid C)} = p(C),
\end{align*}
}
the following derivation for the term $p(M \mid \text{do}(a,y,c))$:
{\small
\begin{align*}
p(C,A,M\mid\text{do}(y)) &= \frac{p(C,A,M,y)}{p(y \mid C,A,M)} {= p(C,A,M)}\\
p(C,M\mid\text{do}(y,a)) &= \frac{p(C,a,M\mid\text{do}(y))}{p(a \mid C,\text{do}(y))} = \frac{p(C,a,M)}{p(a \mid C)} {=p(M \mid a,C) p(C)}\\
p(M\mid\text{do}(y,a,c)) &= \frac{p(c,M\mid\text{do}(y,a))}{p(c \mid \text{do}(y,a))} = \frac{p(M \mid a,c) p(c)}{p(c)} = p(M \mid a,c),
\end{align*}
}
and the following derivation for the term $p(Y \mid \text{do}(a,m,c))$:
{\small
\begin{align*}
p(A,M,Y\mid\text{do}(c)) &= \frac{p(c,A,M,Y)}{p(c)} {=p(A,M,Y \mid c)}\\
p(A,Y\mid\text{do}(c,m)) &= \frac{p(A,m,Y\mid\text{do}(c))}{p(m \mid A,\text{do}(c))} = \frac{p(A,m,Y\mid c)}{p(m \mid A,c)} {= p(Y \mid m,A,c) p(A \mid c)}\\
p(Y \mid \text{do}(c,m,a)) &= \frac{p(a,Y\mid\text{do}(c,m))}{p(a\mid Y,\text{do}(c,m))} = \frac{p(Y \mid m,a,c) p(a \mid c)}{ \frac{ p(Y \mid m,a,c) p(a \mid c) }{ \sum_{\tilde{a}} p(Y \mid m,\tilde{a},c) p(\tilde{a} \mid c) } } =
\sum_{\tilde{a}} p(Y \mid m,\tilde{a},c) p(\tilde{a} \mid c).
\end{align*}
}
We then conclude that $p(Y \mid \text{do}(a))$ is identified from $p(C,A,M,Y)$ via (\ref{eqn:line-4}) and (\ref{eqn:one-district}) by
{\small
\begin{align*}
& \sum_{m,c} p(Y \mid \text{do}(a,m,c)) p(m \mid \text{do}(a,y,c)) p(c \mid \text{do}(a,y,m))
\\
=&
\sum_{m,c} \left( \sum_{\tilde{a}} p(Y \mid m,\tilde{a},c) p(\tilde{a} \mid c) \right) p(m \mid a, c) p(c).
\end{align*}
}
{which recovers the celebrated front-door formula.

\section{Hedges, Corollary 3 and a counterexample}
\label{sec:counterexample}

Whenever the ID algorithm fails to identify a term in (\ref{eqn:line-4}), a particular graphical structure called a \emph{hedge} appears in the graph.  Hedges are defined as follows.

An \emph{$\vec{R}$-rooted C-forest} is a bidirected connected set of vertices $\vec{F}$ in ${\cal G}$ such that $\vec{R} \subseteq \vec{F}$, and there exists a subgraph ${\cal G}^*_{\vec{F}}$ of ${\cal G}$ containing only vertices in $\vec{F}$ and a subset of edges among $\vec{F}$ such that $\vec{F} \subseteq \an_{{\cal G}^*_{\vec{F}}}(\vec{R})$ (every element in $\vec{F}$ has a directed path to an element in $\vec{R}$ in this subgraph).

A \emph{hedge} for $p(\vec{Y} \mid \text{do}(\vec{a}))$ in ${\cal G}(\vec{V})$ is a pair of $\vec{R}$-rooted C-forests $\vec{F},\vec{F}'$ such that $\vec{F} \subset \vec{F}'$, $\vec{F} \cap \vec{A} = \emptyset$, $\vec{A} \cap \vec{F}' \setminus \vec{F} \neq \emptyset$, and $\vec{R}$ is an ancestor of $\vec{Y}$ via directed paths that do not intersect $\vec{A}$.  As Theorem 4 in \cite{shpitser06id} shows, if a hedge for $p(\vec{Y} \mid \text{do}(\vec{a}))$ exists in ${\cal G}(\vec{V})$, $p(\vec{Y} \mid \text{do}(\vec{a}))$ is not identified in (the hidden variable causal model represented by) ${\cal G}(\vec{V})$.

With these preliminaries out of the way, I restate Corollary 3 in \cite{shpitser06id} as follows using notation in this note.

\noindent
{\bf Corollary 3.} \emph{$p(\vec{Y} \mid \text{do}(\vec{a}))$ is identified from $p(\vec{V})$ in (the hidden variable causal model represented by) ${\cal G}(\vec{V})$ if and only if there does not exist a hedge for $p(\vec{Y}' \text{do}(\vec{a}'))$ for any
$\vec{Y}' \subseteq \vec{Y}$, $\vec{A}' \subseteq \vec{A}$.}

A simple counterexample to this claim is shown in Fig.~\ref{fig:example} (c), where we are interested in $p(Y \mid \text{do}(a_1, a_2))$. Note that this interventional distribution is identified, since $\vec{Y}^* = \{ Y \}$, and
$\{ Y \}$ is an intrinsic set, with the valid fixing sequence $\langle A_1, W, A_2 \rangle$.
This yields the following identifying formula:
\[
p(Y \mid \text{do}(a_1, a_2)) = \frac{\sum_{C} p(Y, a_2 \mid a_1, C) p(C)}{ \sum_{C} p(a_2 \mid a_1, C) p(C) }.
\]

However, if we consider $\vec{Y}' = \vec{Y} = \{ Y \}$, and $\vec{A}' = \{ A_2 \}$, we find that $p(\vec{Y}' \mid \text{do}(\vec{a}')) = p(Y \mid \text{do}(a_2))$ is not identified.  Indeed, this query has a hedge structure given by the sets
$\{ Y, W, A_2 \}$ and $\{ Y, W \}$.  Thus, Corollary 3 is not true.

\section{When does the ID algorithm fail?}

Given a vertex set $\vec{S}$ in an ADMG ${\cal G}(\vec{V})$, if $\vec{S}$ is not reachable, there exists a (unique) smallest superset of $\vec{S}$ that is reachable called the \emph{reachable closure} of $\vec{S}$.
I will denote this set by $\langle \vec{S} \rangle_{{\cal G}(\vec{V})}$, following \cite{shpitser18sem}.

The following are three equivalent characterizations of situations when the ID algorithm fails.
\begin{proposition}
Given $p(\vec{Y} \mid \text{do}(\vec{a}))$ and ${\cal G}(\vec{V})$ as inputs, the ID algorithm fails if and only if any one of the following conditions hold:
\begin{itemize}
\item[1] There exists a hedge for $p(\vec{Y} \mid \text{do}(\vec{a}))$ in ${\cal G}(\vec{V})$.
\item[2] Some $\vec{D} \in {\cal D}({\cal G}_{\vec{Y}^*})$ is not intrinsic in ${\cal G}(\vec{V})$.
\item[3] There exists ${\vec D} \in {\cal D}({\cal G}_{\vec{Y}^*})$ such that $\vec{D} \subset \langle \vec{D} \rangle_{{\cal G}(\vec{V})}$.
\end{itemize}
\end{proposition}
In fact, the three conditions above are related.  If $\vec{D} \in {\cal D}({\cal G}_{\vec{Y}^*})$ is not intrinsic, this implies the fixing operator gets ``stuck'' before fixing all elements in $\vec{V} \setminus \vec{D}$.  The set where no further fixing is possible is precisely
$\langle \vec{D} \rangle_{{\cal G}(\vec{V})}$.  Further, it is not difficult to see that ${\cal D}$ and $\langle \vec{D} \rangle_{{\cal G}(\vec{V})}$ are both C-forests rooted in $\{ D \in \vec{D} : \ch_{{\cal G}_{\vec{D}}}(D) = \emptyset \}$, and form a hedge for
$p(\vec{Y} \mid \text{do}(\vec{a}))$ in ${\cal G}(\vec{V})$.

If the ID algorithm does succeed, we obtain the following formula:
\begin{align*}
p(\vec{Y} \mid \text{do}(\vec{a})) = \sum_{\vec{Y}^* \setminus \vec{Y}} \prod_{\vec{D} \in {\cal D}({\cal G}_{\vec{Y}^*})} q_{\vec{D}}(\vec{D} \mid \vec{s}_{\vec{D}}),
\end{align*}
where, as before, $\vec{s}_{\vec{D}}$ are value assignments to $\pa_{\cal G}(\vec{D}) \setminus \vec{D}$ that either consistent with $\vec{a}$, or marginalized over in the outer summation, and each term $q_{\vec{D}}$ is a \emph{Markov kernel} associated with an intrinsic set $\vec{D}$ in ${\cal G}(\vec{V})$, and is a functional of $p(\vec{V})$ given by the repeated application of the fixing operator $\phi$.

Markov kernels associated with intrinsic sets of an ADMG ${\cal G}(\vec{V})$ form the \emph{nested Markov factorization} of an ADMG ${\cal G}(\vec{V})$ \cite{richardson23nested}.  The nested Markov factorization is significant as it captures \emph{all} equality restrictions on the tangent space of the marginal model implied by any hidden variable DAG model associated with the DAG ${\cal G}(\vec{V} \cup \vec{H})$ such that the latent projection of ${\cal G}(\vec{V} \cup \vec{H})$ is ${\cal G}(\vec{V})$ \cite{evans18margins}.
Thus, just as the g-formula may be viewed as a modified Markov factorization of a DAG, the functional output by the ID algorithm for $p(\vec{Y} \mid \text{do}(\vec{a}))$ identified in (the hidden variable causal model represented by) ${\cal G}(\vec{V})$ may be viewed as a modified nested Markov factorization of ${\cal G}(\vec{V})$.

\section*{Acknowledgements}

The author would like to thank Ema Perkovic for bringing a counterexample to the hedge criterion to my attention. 

\small
\bibliographystyle{plain}
\bibliography{references}

\begin{thebibliography}{10}

\bibitem{evans18margins}
Robin~J. Evans.
\newblock Margins of discrete bayesian networks.
\newblock {\em Annals of Statistics}, 46:2623--2656, 2018.

\bibitem{lauritzen96graphical}
Steffen~L. Lauritzen.
\newblock {\em Graphical Models}.
\newblock Oxford, U.K.: Clarendon, 1996.

\bibitem{malinsky19po}
Daniel Malinsky, Ilya Shpitser, and Thomas~S. Richardson.
\newblock A potential outcomes calculus for identifying conditional
  path-specific effects.
\newblock In {\em Proceedings of the 22nd International Conference on
  Artificial Intelligence and Statistics}, 2019.

\bibitem{pearl88probabilistic}
Judea Pearl.
\newblock {\em Probabilistic Reasoning in Intelligent Systems}.
\newblock Morgan and Kaufmann, San Mateo, 1988.

\bibitem{pearl09causality}
Judea Pearl.
\newblock {\em Causality: Models, Reasoning, and Inference}.
\newblock Cambridge University Press, 2 edition, 2009.

\bibitem{richardson23nested}
Thomas~S. Richardson, Robin~J. Evans, James~M. Robins, and Ilya Shpitser.
\newblock Nested {M}arkov properties for acyclic directed mixed graphs.
\newblock {\em Annals of Statistics}, 51(1):334--361, 2023.

\bibitem{robins86new}
James~M. Robins.
\newblock A new approach to causal inference in mortality studies with
  sustained exposure periods -- application to control of the healthy worker
  survivor effect.
\newblock {\em Mathematical Modeling}, 7:1393--1512, 1986.

\bibitem{shpitser18sem}
Ilya Shpitser, Robin~J. Evans, and Thomas~S. Richardson.
\newblock Acyclic linear sems obey the nested markov property.
\newblock In {\em Proceedings of the 34th Annual Conference on {U}ncertainty in
  {A}rtificial {I}ntelligence ({UAI}-18)}, 2018.

\bibitem{shpitser06id}
Ilya Shpitser and Judea Pearl.
\newblock Identification of joint interventional distributions in recursive
  semi-{M}arkovian causal models.
\newblock In {\em Proceedings of the Twenty-First National Conference on
  Artificial Intelligence (AAAI-06)}. AAAI Press, Palo Alto, 2006.

\bibitem{rrs22volume_id}
Ilya Shpitser, Thomas~S Richardson, and James~M Robins.
\newblock Multivariate counterfactual systems and causal graphical models.
\newblock In {\em Probabilistic and Causal Inference: The Works of Judea
  Pearl}, pages 813--852, 2022.

\bibitem{shpitser21proximal}
Ilya Shpitser, Zach Wood-Doughty, and Eric {Tchetgen Tchetgen }.
\newblock The proximal {ID} algorithm.
\newblock \url{https://arxiv.org/abs/2108.06818}, 2021.

\bibitem{tian02on}
Jin Tian and Judea Pearl.
\newblock On the testable implications of causal models with hidden variables.
\newblock In {\em Proceedings of the Eighteenth Conference on Uncertainty in
  Artificial Intelligence (UAI-02)}, volume~18, pages 519--527. AUAI Press,
  Corvallis, Oregon, 2002.

\bibitem{verma90equiv}
Thomas~S. Verma and Judea Pearl.
\newblock Equivalence and synthesis of causal models.
\newblock Technical Report R-150, Department of Computer Science, University of
  California, Los Angeles, 1990.

\end{thebibliography}

\end{document}